\documentclass[aps,prb,twocolumn,showpacs,superscriptaddress,amsmath,amssymb]{revtex4-1}

\usepackage{graphicx}  
\usepackage{bm}

\begin{document}
\title{Electronic structure of vacancy resonant states in graphene: \\a critical review of the single vacancy case}
\author{F. Ducastelle}
\affiliation{Laboratoire d'Etude des Microstructures, ONERA-CNRS, BP 72, 92322 Ch\^atillon Cedex, France}

\begin{abstract}
The resonant behaviour of vacancy states in graphene is well-known but some ambiguities remain concerning in particular the nature of the so-called zero energy modes. Other points are not completely elucidated in the case of low but finite vacancy concentration. In this article we concentrate on the case of vacancies described within the usual tight-binding approximation. More precisely we discuss the case of a single vacancy or of a finite number of vacancies in a finite or infinite system.
\end{abstract}


\maketitle

\section{Introduction}

Vacancies and various point defects in graphene have been the object of numerous studies both theoretical and experimental. It is of course of utmost interest to understand the role of impurities on the electronic structure  of graphene. From the experimental side, let us just mention the problem of doping with nitrogen (or boron) impurities,\cite{Joucken2012,Kondo2012,Zhao2011} the influence of hydrogen adatoms,\cite{Haberer2010,*Haberer2011,Balog2010} or the 
controversial problem of magnetism induced by defects. \cite{Yazyev2010,Palacios2012}

From a theoretical point of view it has been recognized fairly early that the particular electronic structure of graphene with its vanishing density of states at the Dirac points provokes important resonant effects. \cite{Wehling2009} The case of vacancies is particularly interesting since in the simplest tight-binding model with electron-hole symmetry, resonances occur just at these  points. This has been related to the occurrence of so-called zero energy modes where the 
corresponding wave functions are concentrated on the sublattice different from that of the vacancy.\cite{Pereira2008,Kumazaki2008,Toyoda2010,Nanda2012} The case of a single impurity is rather well understood,\cite{Pereira2008,Skrypnyk2011a,Skrypnyk2011b,Farjam2011} but some ambiguities remain\cite{Huang2009} and other points are not completely elucidated in the case of low but finite vacancy concentration, since the usual tools such as the average t-matrix approximation (ATA) or the coherent potential approximation (CPA) are  not accurate enough.\cite{Peres2006,Skrypnyk2006,Pershoguba2009} From the numerical side huge systems (more than 10$^6$ atoms) are necessary to hope to obtain convergence,\cite{Yuan2010,Zhu2012} which is  out of reach of \emph{ab initio} calculations.\cite{Lambin2012}

In this article we concentrate on the case of vacancies described within the usual tight-binding approximation. We first recall this model as well as the properties of the Green functions which will be used at length afterwards. Then, we  discuss the case of a single impurity in a large but finite system. In the presence of a vacancy, one state is substracted from the total density of states. A simple algebraic argument shows on the other hand that a ``quasi-localized" mode 
of zero energy should necessary appear simultaneously. Both effects are clearly associated but we will show that this statement hides some subtleties related to the boundary conditions, which accounts for the apparently different results obtained in the literature.
In the limit of infinite systems, scattering boundary conditions and the  Green function formalism are appropriate, which allows us to discuss in detail the behaviour of the local and total densities of states. 

The case of a finite number of vacancies or impurities can be studied quite similarly, but the thermodynamic limit where the number of impurities is infinite and the concentration $c$ is finite is much more difficult to handle, even in the $c \to 0\,$ limit. This will be discussed elsewhere.

\section{Model and Green functions}
\subsection{The tight-binding model}
We use the simplest tight-binding hamiltonian with single transfer integrals $-t$ connecting the first neighbour $\pi$ orbitals  of the graphene structure: \cite{CastroNeto2009,Katsnelson2012} 
\begin{equation}
H^0 =  -\sum_{\bm{n,m}}{}^\prime \,|\bm{n}\rangle t  \langle \bm{m}|  \; ,
\end{equation}
where $|\bm{n}\rangle$ denotes the $\pi$ state at site $\bm{n}$ and where the prime indicates a sum over first neighbours $\bm{n}$ and $\bm{m}$. The notations used to describe the unit cell and the Brillouin zone are shown in Fig.\ \ref{def} . The corresponding Bloch functions and eigenvalues are:
\begin{figure}[!fht]
\centering
\includegraphics[width=7cm]{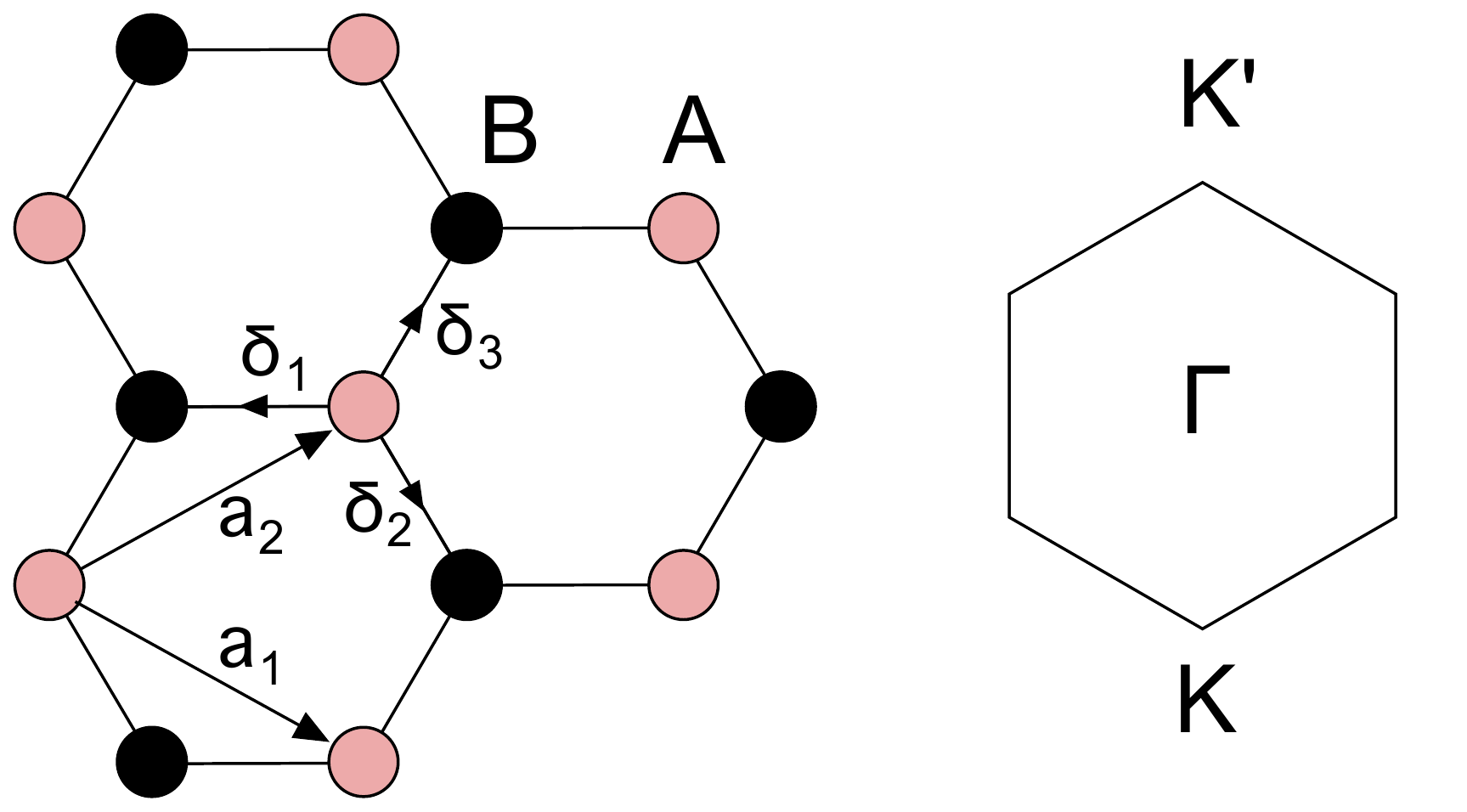}
 \caption{Notations for the graphene lattice and the Brillouin zone used in this article.}
 \label{def}
\end{figure}
\begin{eqnarray}
& &|\bm{k}^{A (B)}\rangle = \frac{1}{\sqrt{N}}\sum_{\bm{n}\in A(B)} e^{i\bm{k.n}}\,|\bm{n}\rangle \nonumber \\
& &E=  \pm t |\gamma({\bm{k}})|\quad;\quad \gamma({\bm{k}})= \sum_{\alpha=1,2,3}e^{i\bm{k.\delta_\alpha}}\; , 
\label{Bloch}
\end{eqnarray}
where $N$ is the number of units cells.
\cite{*[{The number of atoms is therefore equal to $2N$. Notice also that we use a common origin for the two sublattices which will be taken on sublattice $A$. This has advantages and disadvantages as detailed in }] [{.}] Bena2009}
At low energy  $E=\pm\nu q\ ; \  \nu = 3a_{cc}\,t/2$, where $a_{cc}$ is the first neighbour carbon-carbon distance and $\bm{k}=\bm{K}+\bm{q}$ or $\bm{K'}+\bm{q}$.

With our conventions for the $\bm{\delta}$ and $\bm{K}$ vectors, it can be checked that  $\bm{K}_\alpha . \bm{\delta}_\beta = (\beta - \alpha) \frac{2\pi}{3} \mod 2\pi$, and then $\exp(i \bm{K}_\alpha . \bm{\delta}_\beta) = j^{\beta - \alpha}$ where $j= e^{2i\pi/3}$. 

\subsection{Green functions}
\label{Green}
The Green function or resolvent is defined from $G^{\,0}(z)= (z-H^{\,0})^{-1}$, where $z$ is a complex variable, and its matrix elements in the tight-binding basis $G_{\bm{nm}}(z)=\langle \bm{n}|G^{\,0}(z)|\bm{m}\rangle$ are related to the local densities of states  $n_{\bm{nm}}(E)$ through:
\begin{equation}
n_{\bm{nm}}(E) = \langle \bm{n}|\delta(E-H^{\,0})|\bm{m}\rangle= -\lim_{\epsilon\to 0}\frac{\text{Im}}{\pi} \,G^{\,0}_{\bm{nm}}(z)\; .
\end{equation}
Formally $G^{\,0}_{\bm{nm}}(z)$ can be expanded in successive powers of $1/z$. Actually this expansion is convergent for large enough $|z|$\; :
\begin{eqnarray}
G^{\,0}_{\bm{nm}}(z)& = & \langle \bm{n}|\;\frac{1}{z} + \frac{H^{\,0}}{z^2} + \frac{(H^{\,0})^2}{z^3} + \dots |\bm{m}\rangle\nonumber\\
&=& \int dE \; \frac{n_{\bm{nm}}(E)}{z-E} \; . 
\end{eqnarray}
The matrix elements of $(H^{\,0})^{p}$ are related to paths involving $p$ jumps between first neighbours (moments) and are real. 
$G^{\,0}_{\bm{nm}}(z) $ only depends on $\bm{m}-\bm{n}$ and is complex through $z$. As a consequence:
\begin{equation}
G^{\,0}_{\bm{nm}}(z) = G^{\,0}_{\bm{mn}}(z) \quad; \quad (G^{\,0}_{\bm{nm}}(z))^* = G^{\,0}_{\bm{nm}}(z^*) \, .
\label{Gcomplex}
\end{equation}
Furthermore, as in all alternant lattices with first neighbour coupling, we know that in the moment expansion, there are only even or odd paths depending on sites $\bm{n}$ and $\bm{m}$ belonging to the same sublattice or not. As usual we introduce the notations $G^{\,0\,i\,j}_{\bm{n}\bm{m}}, i, j = A, B$ to specify, when necessary, the sublattice $A$ or $B$ of the sites. Finally this implies the following symmetry properties:
\begin{eqnarray}
G^{\,0\,i\,i}_{\bm{n}\bm{m}}(-z)&=& -\,G^{\,0\,i\,i}_{\bm{n}\bm{m}}(z)\nonumber\\
G^{\,0AB}_{\bm{n}\bm{m}}(-z)&=& + G^{\,0AB}_{\bm{n}\bm{m}}(z)\nonumber \; ,
\end{eqnarray}
Close to the real axis, 
\begin{equation}
G^{\,0}_{\bm{nm}}(z=E\pm i\epsilon) = F_{\bm{nm}}(E) \mp i\pi n(E) \quad; \quad \epsilon > 0\; ,
\end{equation}
where  $F_{\bm{nm}}(E)$ is the real part of $G^{\,0}_{\bm{nm}}(E)$, so that using Eq.\ (\ref{Gcomplex}), we obtain:
\begin{eqnarray}
F^{\,0\,i\,i}_{\bm{n}\bm{m}}(-E)&=& -\,F^{\,0\,i\,i}_{\bm{n}\bm{m}}(E) \nonumber\\
F^{\,0AB}_{\bm{n}\bm{m}}(-E)&=& +\, F^{\,0AB}_{\bm{n}\bm{m}}(E)   \; ,
\end{eqnarray}
and opposite relations for the densities of states:
\begin{eqnarray}
n^{\,0\,i\,i}_{\bm{n}\bm{m}}(-E)&=& +\,n^{\,0\,i\,i}_{\bm{n}\bm{m}}(E) \nonumber\\
n^{\,0AB}_{\bm{n}\bm{m}}(-E)&=& -\, n^{\,0AB}_{\bm{n}\bm{m}}(E)   \; ,
\end{eqnarray}
\subsubsection{Explicit expressions}
The subject is well documented,\cite{Wang2006,Bena2009,Bacsi2010,Sherafati2011,Toyoda2010,Nanda2012,Liang2012}
but it is useful here to summarize the results. 
\footnote{The present presentation in particular is quite similar to that published recently  by Nanda \textit{et al.} \cite{Nanda2012} but has been derived independently} 
Explicit expressions for the Green functions are obtained using the Bloch basis:
\begin{eqnarray}
G^{\,0\,ij}_{\bm{n}\bm{m}}&=&\frac{1}{N}\sum_{\bm{k}} e^{i\bm{k.(m-n)}}G^{\,0\,i j}(\bm{k})\ \nonumber \\
G^{\,0\,i j}(\bm{k})&=& \langle\bm{k}^i|G^{\,0} |\bm{k}^j\rangle\; .
\label{green1}
\end{eqnarray}

Hence the familiar $2 \times 2$ matrix representation:
\begin{equation}
G^{\,0}(\bm{k})=\frac{1}{z^2-t^2|\gamma(\bm{k})|^2} \begin{pmatrix} z & -t  \gamma(\bm{k})\\ 
 -t \gamma^*(\bm{k}) & z \end{pmatrix} \; .
\end{equation}
Actually the corresponding integral in Eq.\ (\ref{green1}) can be expressed exactly in terms of elliptic functions and furthermore Horiguchi\cite{Horiguchi1972} has shown that all matrix elements can be obtained in terms of a few of them through recurrence relations. This is not necessarily the best procedure from a numerical point of view\cite{Berciu2010,Sherafati2011} and standard integration or recursion methods are generally more efficient. In practice, since we are principally interested in the  low energy regime, it is very fruitful to obtain simplified expressions which allow us to understand general trends.

\begin{figure}[!fht]
\centering
\includegraphics[width=7cm]{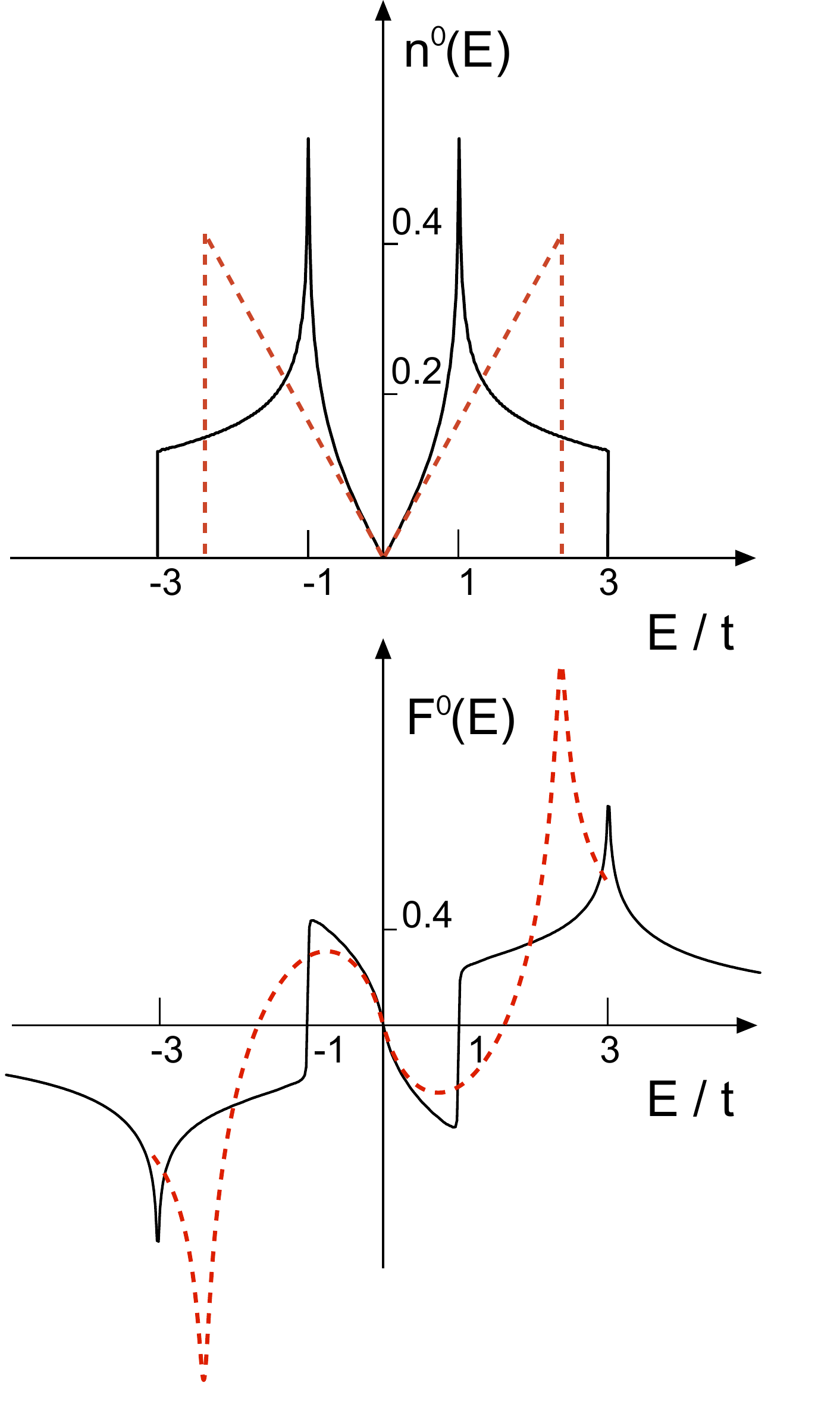}
 \caption{Exact (full line) and approximate (Debye-like approximation, dashed line) diagonal Green fonction: density of states $n^0(E)$ and real part $F^{\,0}(E) \equiv F^{\,0}_{00}(E)$.  The cut-off for the density of states is equal to $W=3^{1/4} \sqrt{\pi} t \simeq 2.33\, t.$}
 \label{g00}
\end{figure}

The diagonal matrix element, independent of $\bm{n}$ plays an important part and is given by:
\begin{eqnarray}
G^{\,0}_{00}(z) &=& \frac{1}{N}\,\sum_{\bm{k}} \frac{z}{z^2-t^2|\gamma(\bm{k})|^2} \nonumber \\
&= &\frac{1}{2N}\,\sum_{\bm{k}} \left( \frac{1}{z-t|
\gamma(\bm{k})|} +  \frac{1}{z+t|\gamma(\bm{k})|}  \right) \, .
\label{green2}
\end{eqnarray}
The imaginary part provides us with the density of states per atom, equal to $n^0(E) \equiv n^0_{00}(E)= |E|/W^2 ; W^2 = \sqrt{3} \pi t^2 $, which is an exact result to first order  in energy. The calculation of the full Green function is more delicate since its real part is the Hilbert transform of the density of states  and involves high energy states where the linear approximation for the dispersion relation is no longer valid. The recipe is known and consists in introducing a 
cut-off $\Lambda$ in $\bm{k}$-space. Equivalently we introduce a cut-off in energy which turns out to be equal to $W$ if we want to keep a normalized density of states. Finally:
\begin{eqnarray}
G^{\,0}_{00}(z) &=& \int dE\,\frac{n(E)}{z-E} \simeq -\frac{z}{W^2} \left.\ln (E^{\,2}-z^2)\right ]_0^W\nonumber \\
& = & \frac{z}{W^2} \ln\frac{z^2}{ (z^2-W^2)} \; ,
\end{eqnarray}
where the proper determination of the logarithm in the complex plane should be taken for Im $z >0$, with a cut along the real axis. The approximation is known to be valid for energies  lower than $t$ (see \textit{e.g.} Ref.\ [\onlinecite{Pershoguba2009}] and Fig.\ \ref{g00}). We will denote it in the following as the Debye-like approximation.
The other matrix elements can be obtained with similar techniques: the imaginary parts can be calculated exactly in the limit $E \to 0$ in terms of Bessel functions, and the full functions depend in principle on a cut-off in $\bm{k}$-space  $\Lambda = W/\nu$ around $\bm{K}$ and $\bm{K'}$. Here however, this cut-off is no longer necessary to achieve convergence of the integrals which finally reduce to Hankel functions in the limit $\Lambda \to \infty$.  For example $G^{\,0AA}_{0 
\bm{n}} (z), \bm{n}\neq 0 $, is obtained from:
\begin{eqnarray}
G_{0\bm{n}}^{AA}(z) &=&  \frac{Sz}{N\pi} \int_0^\Lambda dq \,q \frac{J_0(qn)}{z^2-\nu^2 q^2} \cos \bm{K}.\bm{n} \,\\
&\underset{\Lambda \to\infty}{\simeq}& - i\pi \,\frac{z}{W^2}\, H_0^{(1)}(nz/\nu)\cos \bm{K}.\bm{n}  \nonumber \; .
\label{integral}
\end{eqnarray}
where $S/N$ is the area of the unit cell.
The final result is, when Im$z >0$\,:
\begin{eqnarray}
G^{\,0AA}_{0\bm{n}} (z) &\simeq&\frac{z}{W^2} \ln \frac{z^2}{z^2-W^2} \simeq \frac{z}{W^2} \ln \frac{-z^2}{W^2} \\
G^{\,0AA}_{0\bm{n}} (z) &\simeq&   - i\pi \,\frac{z}{W^2}\, H_0^{(1)}(nz/\nu)\cos \bm{K}.\bm{n}\ \quad n \neq 0 \nonumber \\
G^{\,0AB}_{0 \bm{n}} (z) &\simeq& - i\pi \,\frac{z}{W^2}\, H_1^{(1)}(nz/\nu)  \cos  (\bm{K}.\bm{n}-\omega_{\bm{n}}) \nonumber \; .
\end{eqnarray}
Hankel functions are defined in the upper part of the complex plane as usual, from analytic continuation of their values on the positive  real axis: $H_0^{(1)}(e^{i\pi}z) = -H_0^{(2)}(z); H_1^{(1)}(e^{i\pi}z) = H_1^{(2)}(z)$. For example $H_0^{(1)}(x) = J_0(x) + i\, Y_0(x)$ when $x >0$ becomes $- J_0(x) + i\, Y_0(x)$ when $E<0$. When Im$z <0, H_{0,1}^{(1)}(z)$ should be replaced by $-H_{0,1}^{(2)}(z)$; $\omega_{\bm{n}}$ is the angle between $\bm{n}$ and the $Ox$ axis which is, according to our convention, the armchair direction. Notice that this precise form of the argument 
of the cosine depends on our choice of the $\bm{K}$ point because in the case of the $G^{\,0}_{AB}$ Green function, $\bm{n}$ is not a lattice vector.

\subsubsection{Low energy limit  $E \to 0$}
Since the Bessel functions depend on the product $E\,n$,  the $E \to 0$ and $n \to \infty$ limits do not commute. We consider here the $E \to 0$ limit. Using the standard properties of  Bessel functions in the limit $x \to 0$ :
\begin{eqnarray*}
J_0(x) &\simeq& 1-\frac{x^2}{2} \quad ; \quad Y_0(x) \simeq \frac{2}{\pi} (\ln \frac{x}{2} + \gamma ) \\
J_1(x) &\simeq & \frac{x}{2} \quad ; \quad Y_1(x) \simeq -\frac{2}{\pi x} \; ,
\end{eqnarray*}
where $\gamma$ is the Euler constant, we obtain:
\begin{eqnarray}
G^{\,0AA}_{00}(E)& \simeq &\left[\frac{2E}{W^2} \ln \frac{|E|}{W}  -i\pi \frac{|E|}{W^2}\right]  \\
G^{\,0AA}_{0\bm{n}} (E) & \simeq &\left[\frac{2E}{W^2} (\ln \frac{|E|n}{2\nu} + \gamma) -i\pi \frac{|E|}{W^2}\right] \cos \bm{K}.\bm{n} \nonumber\\
G_{0\bm{n}}^{\,0AB}(E) &\simeq &\left[ \frac{-2\nu}{nW^2} - i\pi \frac{n}{2\nu} \,\frac{\text{sgn}\,(E) E^2}{W^2}\right]  
\cos  (\bm{K}.\bm{n}-\omega_{\bm{n}}) \nonumber \; .
\label{limE0}
\end{eqnarray}
Notice finally that one cannot simply set $n=0$ in $G^{\,0AA}_{0\bm{n}} (z) $ to obtain the diagonal Green fonction. The  high energy cut-off corresponds to a low distance cut-off about 0.7 $a_{cc}$. All these results are in full agreement with previous estimates.\cite{Nanda2012}
The behaviour of the Green functions at the origin is therefore different from that at the usual van Hove singularities (discontinuity or logarithmic singularity). %
\footnote{This does not agree with Horiguchi's assertion\cite{Horiguchi1972} that only the first neighbour Green function $G^{0AB}_{0\bm{n}}$ is continuous at the origin}
In particular all $G^{\,0AA}_{0\bm{n}}(E)$ vanish, whereas all $G^{\,0AB}_{0\bm{n}}(E)$ behave as $\cos  (\bm{K}.\bm{n}-\omega_{\bm{n}})/n$ and are finite, because the cosine never vanishes (see below). 

\subsubsection{Large distance limit  $n \to \infty$}
The large distance behaviour is governed by the Hankel functions:
\begin{equation}
 G^{\,0AA}_{0\bm{n}}(E)  \underset{n\to \infty}\sim    -\frac{i}{W}\sqrt{\frac{a}{n}\frac{E}{t} } \exp \left[i\left(\frac{nE}{\nu}-\frac{\pi}{4}\right)\right ] \cos  \bm{K}.\bm{n} 
 \end{equation}
 \begin{equation} 
 \begin{split}
G^{\,0AB}_{0 \bm{n}}& (E) \underset{n\to \infty}\sim   \\
 &-\frac{i}{W}\sqrt{\frac{a}{n}\frac{E}{t} } \exp \left[i\left(\frac{nE}{\nu}-\frac{3\pi}{4}\right)\right ]  \cos  (\bm{K}.\bm{n}-\omega_{\bm{n}})  \; ,
 \end{split} 
 \end{equation}
where here $a=\sqrt{3}a_{cc}$ is the length of  unit cell vectors $\bm{a}_{1,2}$.
In fact this limit can be obtained directly using stationary phase arguments\cite{Koster1954}  and is practically exact at low --- but finite --- energy.

\subsubsection{Symmetry properties}
The approximate expressions for the Green functions have the advantage to factorize the dependence on distance. The Bessel functions depend only on the modulus $n$ of $\bm{n}$ whereas the spatial symmetry is governed by the cosine factors. 

\begin{figure}[!fht]
\centering
\includegraphics[width=6cm]{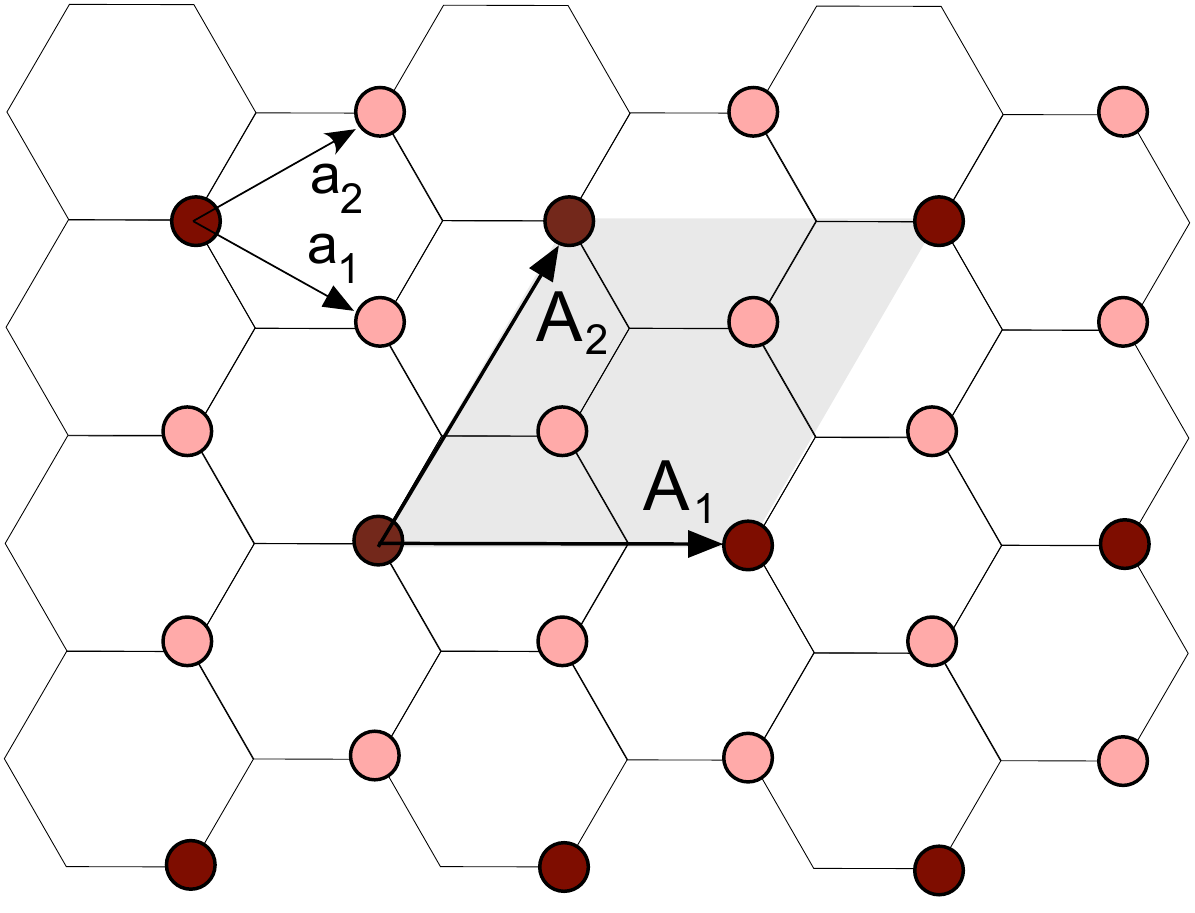}
\caption{$\sqrt{3}\times\sqrt{3}$ superstructure. The factor $\cos  \bm{K}.\bm{n}$ involved in $G^{\,0AA}_{0\bm{n}}$  takes value 1 and -1/2 on the superlattice (dark atoms) and on the two other atoms of the motif (light atoms), respectively.}
\label{R3}
\end{figure}

\begin{figure}[!fht]
\centering
\includegraphics[width=6cm]{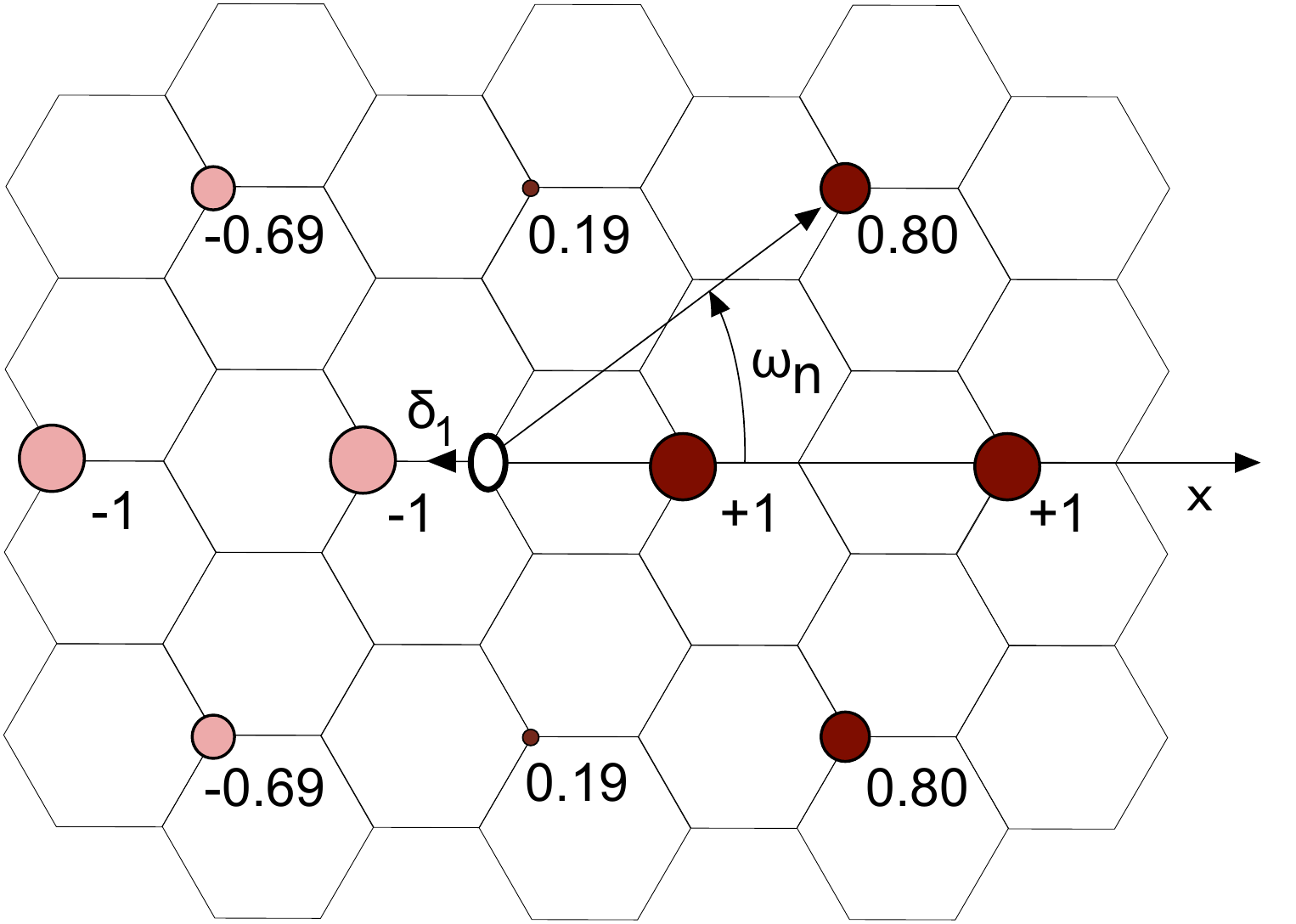}
\caption{The geometrical factor involved in $G^{\,0AB}_{0\bm{n}}$ is equal to $\cos \omega_{\bm{n}}$ for the atoms shifted from the  $\sqrt{3}\times\sqrt{3}$ superstructure by the vector $\bm{\delta}_1$. Its value is indicated close to these atoms. The size of the atoms is approximately proportional to the absolute value of this factor.}
\label{GAB}
\end{figure}

\begin{figure}[!fht]
\centering
\includegraphics[width=5cm]{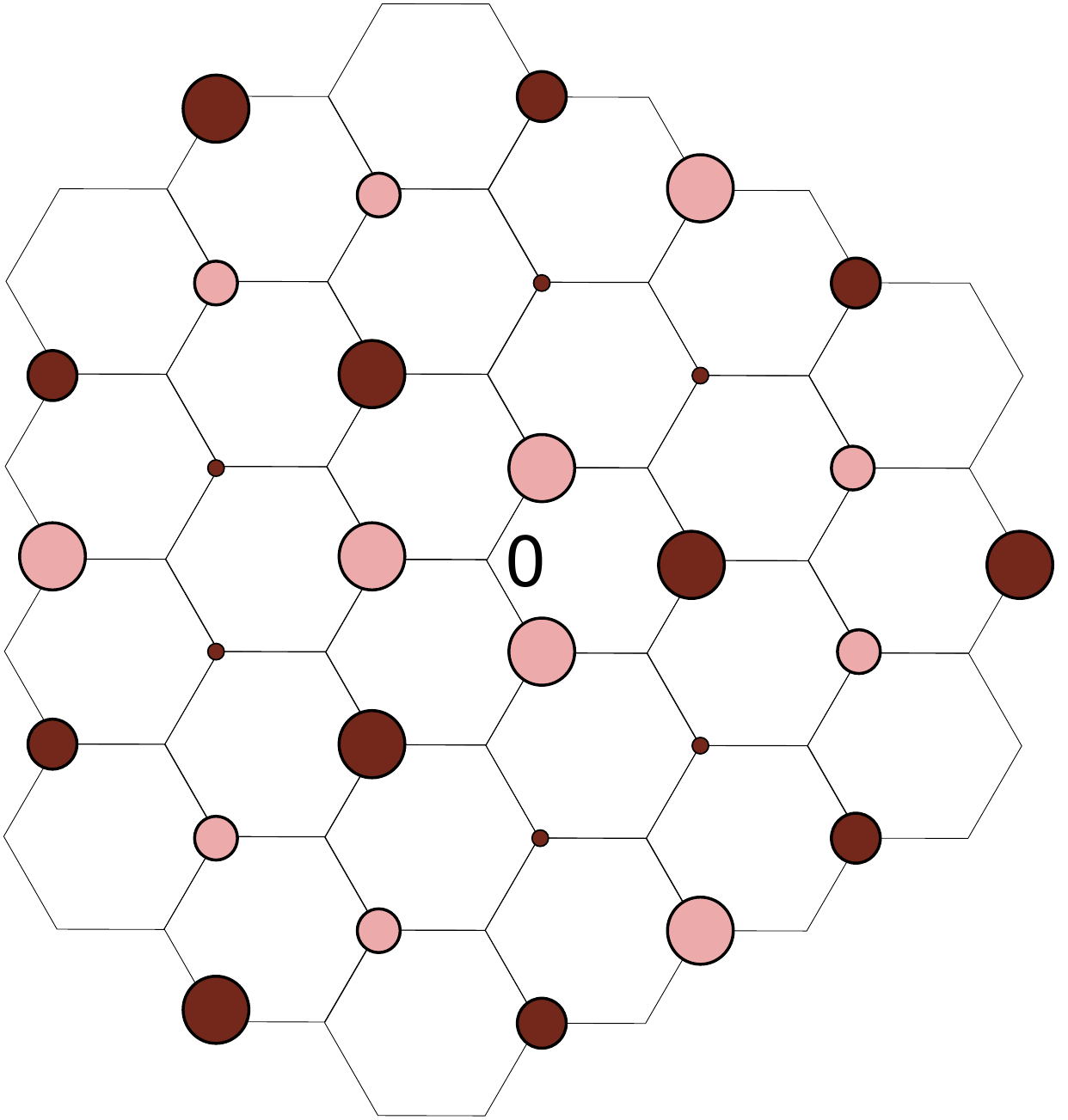}
\caption{Qualitative map of the full  geometrical factor involved in $G^{\,0AB}_{0\bm{n}}$.}
\label{GAB++}
\end{figure}

\paragraph{$G^{\,0AA}$ Green functions. }  For $G^{\,0AA}_{0\bm{n}}$, this factor is equal to $\cos  \bm{K}.\bm{n}$ which has the symmetry of the so-called $\sqrt{3}\times\sqrt{3}$ superstructure. It takes values 1 and $-1/2$ depending on the sites belonging to the superlattice or to its motif, respectively. Globally, $G^{\,0AA}_{0\bm{n}}$ has the full C$_{6v}$ point symmetry of the hexagonal lattice (Fig.\  \ref{R3}).
\paragraph{$G^{\,0AB}$ Green functions. }  The case of the Green functions connecting different $A$ and $B$ sublattices is more subtle. Any vector $\bm{n}$ in this case can be decomposed into a vector of the $\sqrt{3}\times\sqrt{3}$ structure and a vector of type $\bm{\delta}_\alpha$. Consider first the atoms obtained from $\bm{\delta}_1$ along $Ox$ in the negative direction. For these atoms $\bm{K}.\bm{n} = 0 \mod{2\pi}$ and the geometrical factor is simply  $\cos \omega_{\bm{n}}$ which takes values $\pm 1$ on the $0x$ axis. Its absolute value decreases when deviating from this axis. In the zig-zag direction the angle is never equal to $\pi/2$ but reaches this value asymptotically (Fig.\ \ref{GAB}).
The other atoms are simply obtained by applying a three-fold symmetry about the origin, which yields the map shown in Fig.\ \ref{GAB++}.

It is interesting to notice that all atoms in the armchair direction along $Ox$ are already obtained with the single family shown in Fig.\ \ref{GAB}
so that $|\text{Re}\,G^{\,0AB}_{0\bm{n}}(E)| \sim 1/n$ in this case without any oscillatory behaviour at low energy. On the other hand, in the $Oy$ zig-zag direction the Green function is very weak for the family of sites shown in Fig.\ \ref{GAB}  contrary to the two other ones obtained from $\pm 2\pi/3$ rotations. This has been studied in more detail in Refs [\onlinecite{Nanda2012}] and [\onlinecite{Liang2012}].

\subsubsection{Discussion}
\begin{figure}[!fht]
\centering
\includegraphics[width=7cm]{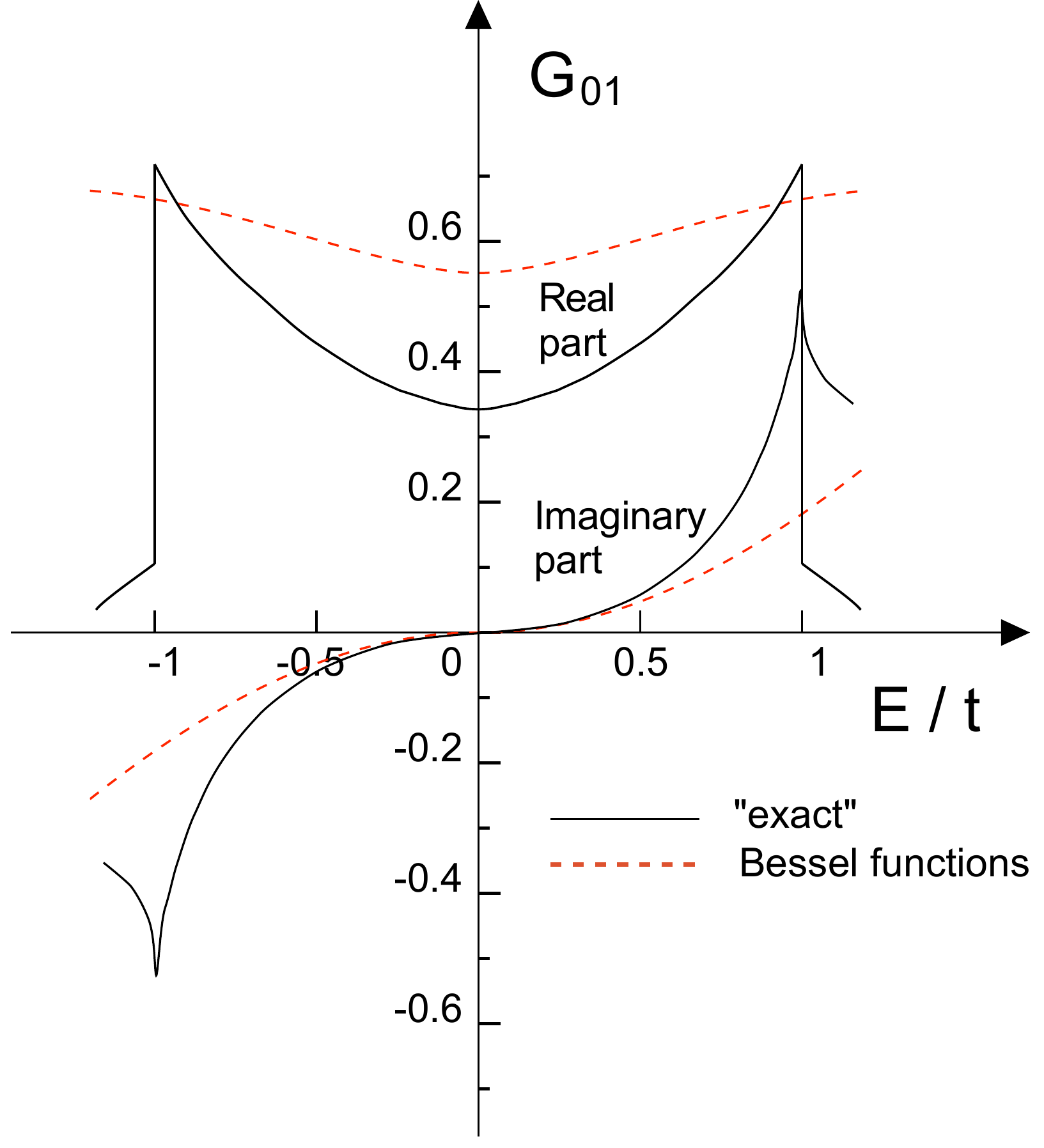}
\caption{Green function for the first neighbours: comparison between recursion calculations and the Bessel function approximation.}
\label{G01}
\end{figure}

\begin{figure}[!fht]
\centering
\includegraphics[width=6cm]{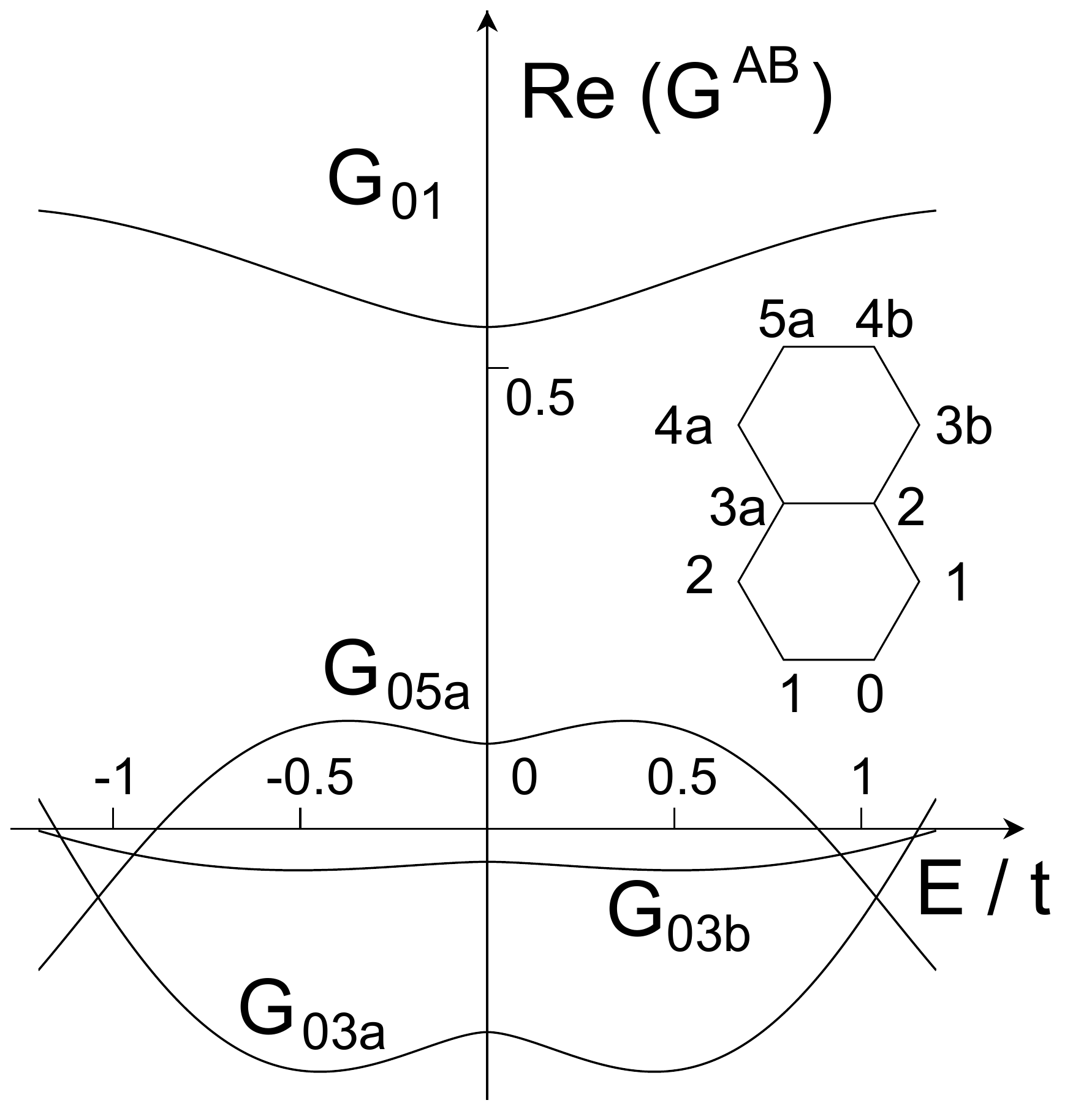}
\caption{Real part of the first $AB$ Green functions. The neighbours are labelled by the number of first neighbour steps; a further label: $a,b, \dots $ corresponds to increasing distances.}
\label{ReGAB}
\end{figure}

As pointed above, an important result is that the only non-vanishing Green functions when $E \to 0$ are the real parts of the off-diagonal matrix elements connecting the two different $A$ and $B$ sublattices. Since the real parts are less accurate than the imaginary parts, it is useful to test the accuracy of the previous formul\ae. Precisely at $E=0$ it is in fact possible to account analytically for the cut-off $W$ with the result:
\begin{equation}
\text{Re}\,G^{\,0AB}_{0\bm{n}}(E) \underset{E\to 0} \simeq\frac{-2\nu}{nW^2} [1-J_0(nW/\nu)] \cos  (\bm{K}.\bm{n}-\omega_{\bm{n}}).
\end{equation}
Since $W/\nu \simeq 1.56/a_{cc}$ the correction decreases rapidly with $n$ but is non negligible for the first neighbours for which $n=a_{cc}$.
Let $G^{\,0}_ {01}$ denotes the first neighbour Green function, the previous formula gives $G^{\,0}_ {01}(E=0) \simeq 0.29 /t$ whereas the asymptotic formula for an infinite cut-off gives $0.55 /t$. The exact result is $1/3t$ and can be deduced from the exact relations obtained from the identity $(z-H^{\,0})\,G^{\,0}=1$. Taking the diagonal matrix elements of both sides yields indeed: $G^{\,0}_ {01} = (1-zG^0_{00})/3t = 1/3t$ when $z=0$. In Fig.\ \ref{G01}  we compare the 
``exact'' result obtained from the recursion method with the Bessel function approximation; the result is qualitatively good if not quantitatively, but would be improved using a cut-off  as mentioned above (see also Ref.\ [\onlinecite{Wang2006}]).
\footnote{Notice also that with our choice of a negative transfer integral the real part of $G^{\,0}_ {01}$ should be positive.}
Taking then the matrix elements between first neighbours we obtain also an exact expression for the second neighbour Green function $G^{\,0}_ {02} = [(z^2-3t^2)G^0_{00} - z]/6t^2$. In Fig.\ \ref{ReGAB} we show the real parts of the first $G^{\,0AB}_{0\bm{n}}$ Green functions using Bessel functions. By comparison with exact numerical calculations, it can be checked that the larger $n$, the better the accuracy of the Bessel function approximation.

It should be pointed out also that at finite energy,  the Bessel functions begin to oscillate as soon as $nE/\nu$  is  larger than about 2. If energies are measured in units of $t$ and distances in units of $a_{cc}$, this means that the low energy expansion of the Green functions $G^{\,0}_{0\bm{n}}(E)$ are valid when $E < 1/n$.

\section{Electronic strcture of the vacancy}
\subsection{Zero modes or not ?}
The existence of particular states of zero energy, the so-called zero modes, has been predicted using fairly simple algebraic arguments which apply to alternant lattices described within a first neighbour tight-binding approximation.\cite{Pereira2008} The vacancy here is modelled by removing the corresponding state at the vacancy site taken at the origin ($A$ sublattice). Although the single vacancy is known to display a slight reconstruction\cite{Elbarbary2003,Amara2007} the 
model remains useful and applies to other situations (hydrogen adatom, pyridine-like  nitrogen defect). Removing an atom can be made either explicitly by introducing the hamiltonian $H$ in the new basis:
\begin{equation}
H = Q^0H^{\,0}Q^0 \quad; \quad Q^0=1-P^0 \equiv 1- | 0\rangle \langle 0 | \;,
\end{equation}
or by introducing a local potential $V=| 0\rangle U \langle 0 |$ and taking the limit $U \to\infty$, which forces the wave function to vanish on site $0$:
\begin{equation}
H = H^{\,0}+| 0\rangle U \langle 0 | \; .
\end{equation}
The only difference is that in the latter case, one bound state is repelled to infinity. Otherwise the two approaches are equivalent, but the latter is known to be more easy to handle. It will be clear in the context what precise hamiltonian is used.

Now, starting from pristine graphene with an equal number $N$ of $A$ and $B$ sites, the energy spectrum is symmetric about the origin. In the presence of a vacancy the hamiltonian applies the space of $B$ states of dimension $N$ on the space of $A$ states of dimension $N-1$, and there is necessarily a non trivial linear combinaison of $B$ states whose image vanishes. This supplementary state is supposed to be at the origin of the resonance found in the 
density of states. Actually it is always dangerous to apply theorems valid for finite systems to infinite systems or systems with special boundary conditions, and this is exactly what happens here as we show now.

\subsubsection{Vacancy in a finite system}
\begin{figure}[!fht]
\centering
\includegraphics[width=5cm]{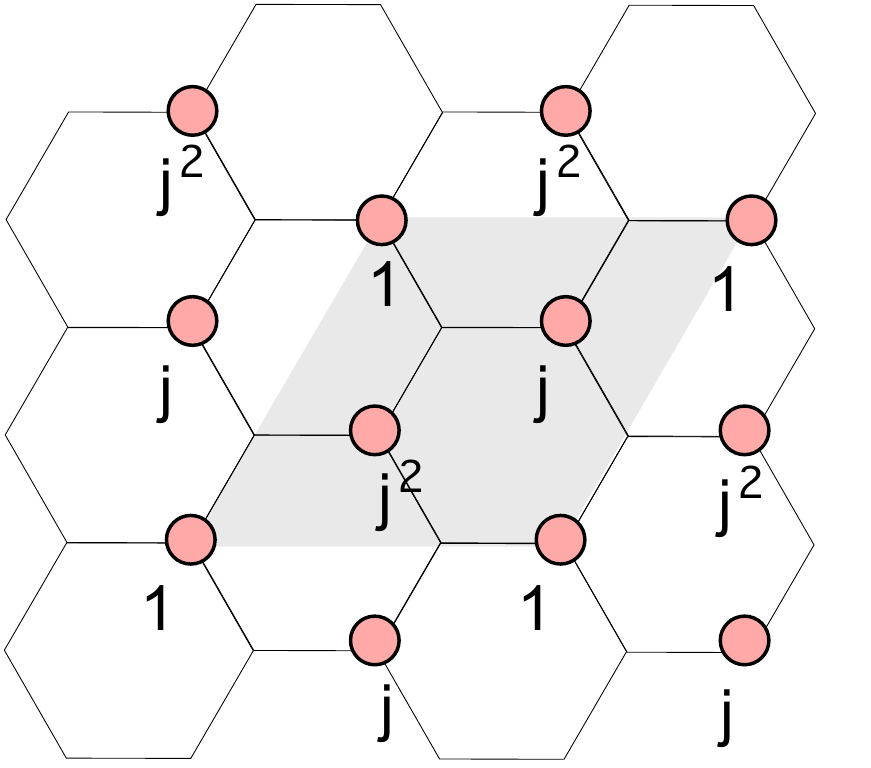}
\caption{Bloch state of zero energy. There are three other possibilities, first by taking the complex conjugate and then by considering the other sublattice.These states have the periodicity of the $\sqrt{3}\times\sqrt{3}$ structure.}
\label{BlochR3}
\end{figure}

The simplest way to build a perfect finite graphene structure is to introduce periodic boundary conditions. This is possible in many different ways as we have learnt in the case of carbon nanotubes where any non trivial lattice vector can define a wrapping vector $\bm{C} = p \bm{a}_1 + q \bm{a}_2$ where $p$ and $q$ are integers. A second independent vector  defines then a torus. Nanotubes are metallic or semi-conductor depending on $p-q$ being a multiple of 3 
or not. This depends on $\bm{K}$ being an allowed wave vector for the Bloch functions or not. This can be extended to a torus. For simplicity we will consider $p \times p$ tori in the following. Then $\bm{K}$ is  an allowed wave vector if $p$ is a multiple of 3. It is instructive to look more precisely at the Bloch fonctions $| \psi \rangle$ at the  $\bm{K}$ and $\bm{K'}$ points. We have four possibilities: $| \bm{K}^{A (B)}\rangle$ and $| \bm{K'}^{A (B)}\rangle$. The 
corresponding values of the amplitudes  $\langle \bm{n}|\psi \rangle$ are the cubic roots of unity, $1, j=\exp(2i\pi/3)$ and  $j^2=\exp(-2i\pi/3)$ which gives four possibilities indeed. A given state has its amplitudes on a single sublattice and amplitudes rotating with the sequence $(1,j,j^2)$ or $(1,j^2,j)$ on every triangle of the sublattices (Fig.\ \ref{BlochR3}).
In real space the Bloch states live therefore on the $\sqrt{3}\times\sqrt{3}$ structure. The wrapping vectors must therefore be consistent with this periodicity and we recover the fact that $p$ should be a multiple of 3. 

\paragraph{Metallic case}
To summarize, in the case of metallic tori, there are exactly four states of zero energy. Otherwise (semiconductor tori) there is no state of vanishing energy. This contradicts the argument sometimes put forward that zero modes only appear in the case  where the numbers of $A$ and $B$ sites are different.

What happens then in the presence of a vacancy?  Of course in the case of a metallic torus, states on the $B$ sublattice are not perturbed and remain at zero energy. But there is another possibility frequently forgotten, which is to consider the linear combination $| \bm{K}^A\rangle - | \bm{K'}^A\rangle$, which also vanishes on the vacancy site. Therefore three states survive at zero energy. And there is no other possibility. 

In the presence of a localized potential actually,  the tight-binding problem is basically exactly solvable for finite as well as for infinite systems. \cite{KosterSlater1954,Lifshitz1965} The eigenenergies are  obtained from the equation:
\begin{equation}
U G^{\,0}_{00}(E) = 1 \;,
\label{SK}
\end{equation}
where here $G^{\,0}_{00}(E)$ is the real Green function for the finite system considered:
\begin{equation}
G^{\,0}_{00}(E) = \sum_\alpha \; \frac{|\langle 0| \alpha\rangle|^2}{E-\varepsilon_\alpha} \; ,
\end{equation}
where the $|\alpha\rangle$ and $\varepsilon_\alpha$ are the (discrete) eigenstates and eigenenergies of the unperturbed system.
In the present case the unperturbed eigenstates are Bloch functions and all matrix elements $\langle 0| \alpha\rangle$ are equal to $1/\sqrt{N}$ and the solution of Eq.\ (\ref{SK}) is obtained through the usual procedure of finding the intersections of $G^{\,0}_{00}(E)$ with $1/U$. Because of the divergences at each $\varepsilon_\alpha$ the new eigenvalues are found to alternate with the unperturbed ones. There are problems  in the case of degeneracies but then 
it is sufficient to imagine that weak appropriate perturbations lift the degeneracies, so that a level with degeneracy $g$ will give rise to $g-1$ levels at the same energy, as we found above. Otherwise, the levels alternate strictly. In the case of the vacancy, $U \to\infty$ and the new eigenvalues are the zeros of  $G^{\,0}_{00}(E)$. The conclusion is that in the case of metallic tori, there are three and just three \textit{delocalized} states of zero energy.

\paragraph{Semiconductor case}
The case of semiconductor tori is different. The $\bm{K}$ points are no longer allowed and there is a small but finite gap of order $1/L$ where $L$ is the linear size of the system. By symmetry however $G^{\,0}_{00}(E=0) =0$ so that there is a single zero mode in that case. Its wave function has been derived by Pereira \textit{et al.}\cite{Pereira2006} by combining localized edge states. There is actually a direct and simple proof (see also Refs.\ 
[\onlinecite{Kumazaki2008,Nanda2012}]). Consider the state $G^{\,0} (z=0)|0\rangle$ and let us apply the vacancy hamiltonian $H=Q^0H^{\,0}Q^0$ on this state, using the fact that, by definition  $H^{\,0}G^{\,0} (z=0) = -1$. We have:

\begin{eqnarray}
H G^{\,0} (z=0)|0\rangle &= &Q^{\,0}H^{\,0}Q^{\,0} G^{\,0} (z=0)|0\rangle\\
& = &-Q^{\,0} |0\rangle - Q^{\,0}H^{\,0}P^{\,0}G^{\,0} |0\rangle =0  \; . \nonumber
\end{eqnarray}

The first term in the right hand side vanishes indeed by definition. As for the second one,  it is proportional to $G^{\,0}_{00}(E)$ which vanishes also when $E=0$. The origin being on a $A$ sublattice, the considered state vanishes on this sublattice because of the symmetry properties of the Green functions. The components $\psi_{\bm{n}} =\langle \bm{n}|\psi \rangle$ of this state are just the matrix elements of the bare Green function $G^{\,0}_{0\bm{n}}(E=0)$. For a 
large system the sum defining it can be safely replaced by an integral. We can use the results of the previous section so that:
\begin{equation}
\psi_{\bm{n}} = G^{\,0}_{0\bm{n}}(E=0)\sim \frac{1}{n}  \cos  (\bm{K}.\bm{n}-\omega_{\bm{n}}) \; ,
\end{equation}
when $\bm{n}$ belongs to the $B$ sublattice, and vanishes on the $A$ sublattice. In the continuous limit, this becomes $\psi({\bm{r}}) \sim e^{i\bm{K.r}}/(x+iy)$ + complex conjugate, which is in complete agreement with previous estimates if appropriate correspondences for the coordinates and the $\bm{K}$ points are made.\cite{Pereira2006,Pereira2008} To normalize this state, we calculate $\sum_{\bm{n}} |\psi_{\bm{n}} |^2 = \langle0|(G^{\,0})^2 |0\rangle =-dG^{\,0}
_{00}(E)/dE$ for $E=0$, which is of order $\ln N$ for a (finite) large system, as noticed by many authors. In the present case this can be shown using $\langle0|(G^{\,0})^2 |0\rangle \sim \sum_\alpha (1/\varepsilon_\alpha^2)  \sim\int_{1/L}^\Lambda qdq/q^2 \sim \ln N$.

\paragraph{Discussion}
One conclusion of our discussion is that there is well defined quasi localized zero mode associated to the vacancy if the initial graphene system has no zero mode already, and this depends on the precise boundary conditions used. Quasi localized and itinerant states as well as more or less hybridized states can coexist close to zero energy. This accounts for some previous puzzling results. For example  Pereira \textit{et al.}.\cite{Pereira2008} reported a first localized state 
at zero energy and a second state completely itinerant, which corresponds to our ``semiconductor'' case. Conversely, Huang \textit{et al.}. used ``metallic'' boundary conditions and did found itinerant zero modes.\cite{Huang2009} Evidently, there are cases where a genuine gap opens: nanoribbons,\cite{Palacios2008} semiconducting nanotubes, in which case the occurence of a true bound state is clear.

\subsubsection{Finite number of vacancies in a finite system}
We consider here semiconducting tori. Since the vacancy state $G^{\,0}(E=0)|0\rangle$ has no amplitude on the $A$ sites, it is not sensitive to other vacancies on this sublattice. Finally with a finite set of vacancies at sites $\bm{n}_v$ is associated a finite number of zero modes $G^{\,0}(E=0)|\bm{n}_v\rangle$. This can also be seen by treating the case of a second vacancy at site $A2$ as a perturbation of the case of a single vacancy at site $A1$. We have again to 
solve an equation of type $\sum_\alpha \; |\langle A2| \alpha\rangle|^2/(E-\varepsilon_\alpha) = 0$ where now $\alpha$ labels states in the presence of the vacancy $A1$. There is then a state of zero energy, but its weight $|\langle A2| \alpha\rangle|^2$ vanishes so that this function has still a zero (and not a divergence) at the origin. If instead the second vacancy is on a $B$ site the weight no longer vanishes and the zero mode disappears.

\subsubsection{ Vacancy in an infinite system}
\begin{figure}[!fht]
\centering
\includegraphics[width=6
cm]{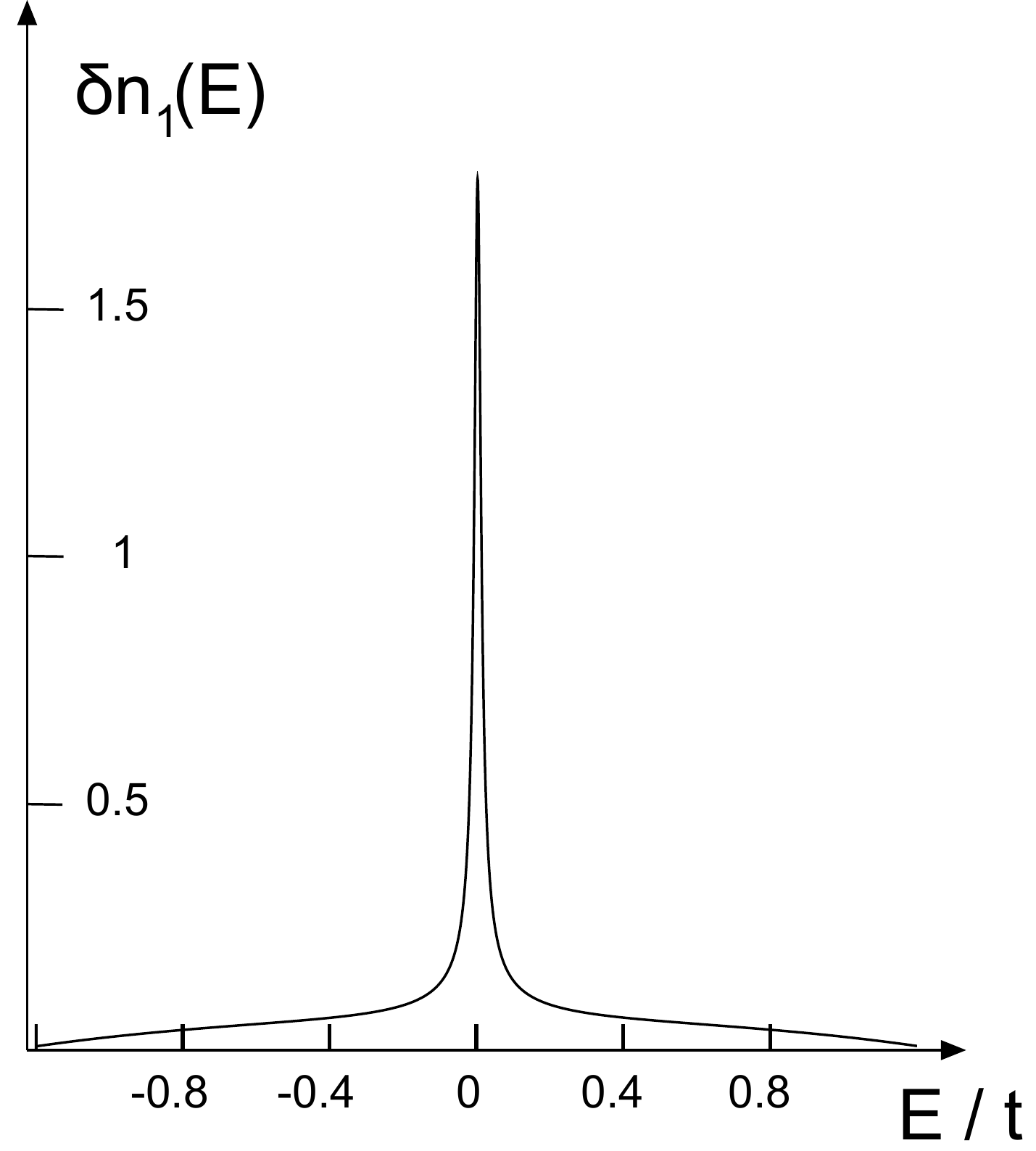}
\caption{Variation of the local density of states on the first neighbours of the vacancy. The shape is similar for all $B$ neighbouring sites.} 
\label{n1}
\end{figure}

\begin{figure}[!fht]
\centering
\includegraphics[width=6
cm]{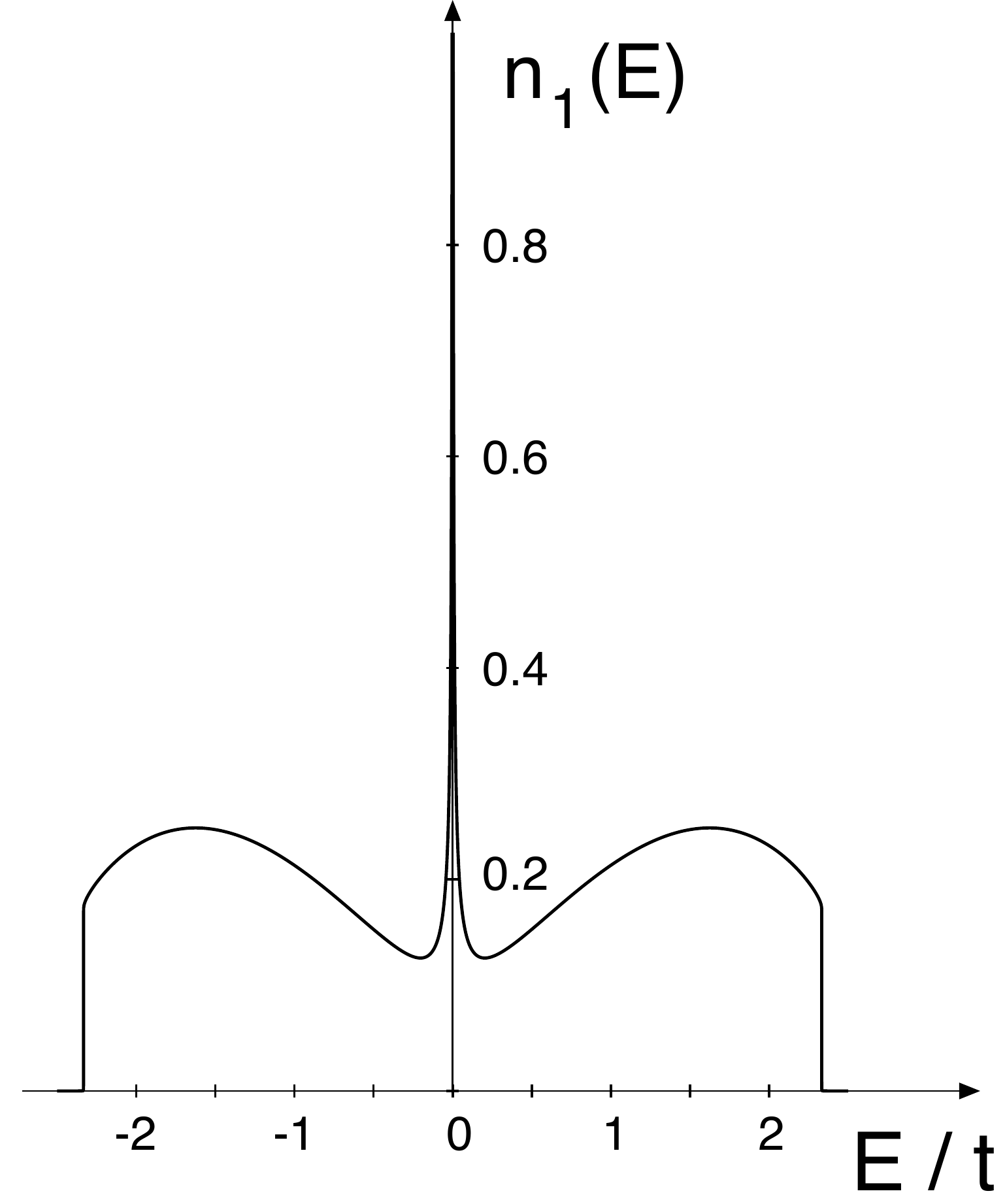}
\caption{Total density of states on the first neighbours of the vacancy (Debye-like approximation). Compare with the unperturbed density of states (dashed lines) in Fig.\ \ref{g00}.}
\label{fulln1}
\end{figure}

\begin{figure}[!fht]
\centering
\includegraphics[width=6cm]{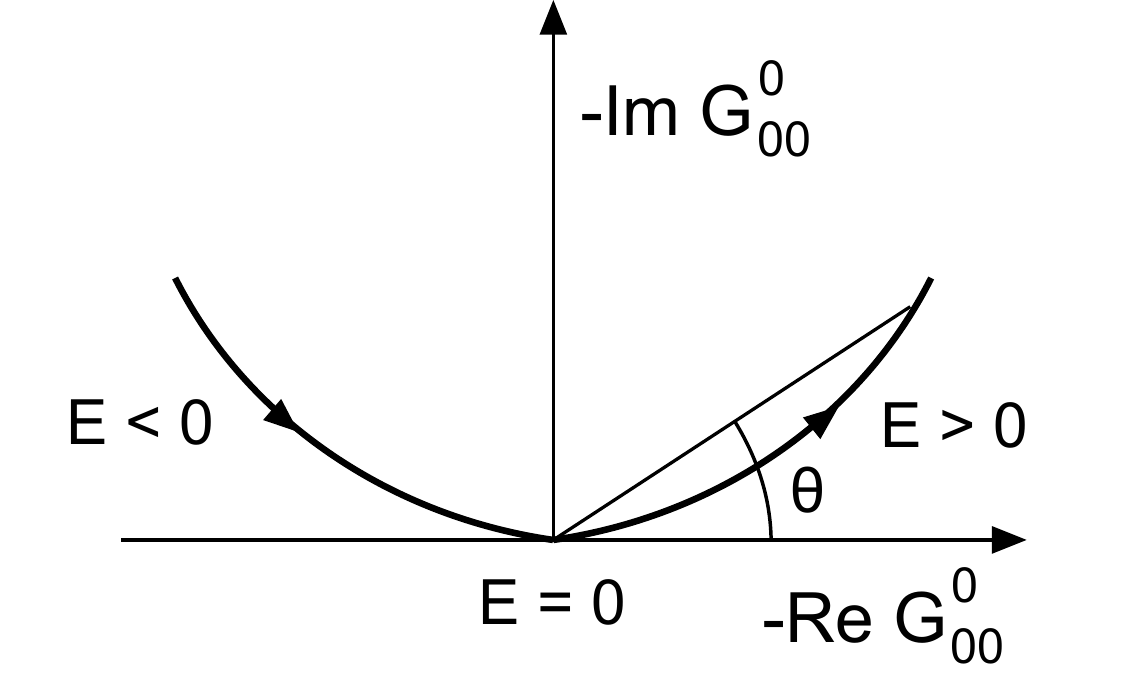}
\caption{Trajectory of the diagonal Green function in the complex plane when $z=E+i\varepsilon$ moves close to $E=0.$ The angle $\theta$ jumps from $\pi$ to 0 on an interval of width about $\varepsilon$.}
\label{arg}
\end{figure}

From the above discussion it is clear that it is preferable to use a formalism that does not depend on the boundary conditions. The usual method for that introduced in Section \ref{Green} is to add imaginary parts $E \rightarrow E+i\varepsilon$ and calculate density of states variations by taking the limit $\varepsilon \to 0$ after taking the limit $N \to \infty$. In the case of a localized potential the exact expression of the Green function is known:
\begin{eqnarray}
G_{\bm{nm}} &=& G^{\,0}_{\bm{nm}} + G^{\,0}_{\bm{n}0} \; \tau \;G^{\,0}_{0\bm{m}}\nonumber \\
\tau &=& U/(1-G^{\,0}_{00}U)  \; ,
\end{eqnarray}
where $\tau$ is the t-matrix.
In the case of a vacancy, in the limit $U \to\infty$ this reduces to:
\begin{equation}
G_{\bm{nm}} = G^{\,0}_{\bm{nm}} -G^{\,0}_{\bm{n}0} \;G^{\,0}_{0\bm{m}}/G^{\,0}_{00}  \; .
\label{Goff}
\end{equation}
The variation $\delta n_{\bm{n}}(E)$ of the local density of states at site $\bm{n}$ is then given by:
\begin{eqnarray}
\delta n_{\bm{n}}(E) &=&  -\lim_{\epsilon\to 0}\frac{\text{Im}}{\pi} \,\delta G_{\bm{nn}} \nonumber\\
\delta G_{\bm{nn}}&=& -(G^{\,0}_{0\bm{n}})^2/G^{\,0}_{00}  \; .
\end{eqnarray}
Using the results of Section \ref{Green} for the Green functions , $\delta G_{\bm{nn}}$ is then found to vanish at $E=0$  on $A$ sites whereas it diverges  on the B sites. For these sites the variation of the density of states is given by:
\begin{equation}
\delta n_{\bm{n}}(E) \simeq (G^{\,0AB}_{0\bm{n}}(E=0))^2  W^2\frac{1}{|E|(\ln (W^2/E^2-1))^2}   \; ,
\label{eldens}                  
\end{equation}
and is shown in Fig.\ \ref{n1} in the case of first neighbours, $n=a_{cc}$. The full density of states $n_1(E)$ is shown in Fig.\ \ref{fulln1}. The number of states between $E=-t$ and $E=+t$ is about 0.2, but is only equal to 0.08 when the integration is made between -0.1 and 0.1. These states are substracted from the wings of the unperturbed density of states. The corresponding $\delta n_{\bm{n}}(E)$ for $A$ sites is quasi negligible, of the order of 0.05 $t^{-1}$ for the second neighbours.
Up to a multiplicative factor the curves are similar for all $B$ sites, for the close neighbours at least. At larger distances the oscillations due to $G^{\,0AB}_{0\bm{n}}$ appear. As a result interferences occur when summing all contributions. We have actually an exact formula for the total variation:
\begin{equation}
\delta n(E) = \sum_{\bm{n}}\delta n_{\bm{n}}(E) = -\frac{\text{Im}}{\pi} \,   \frac{d}{dE} \ln  G^{\,0}_{\bm{00}} \; .
\end{equation}
For the integrated density of states $N(E) = \int^E dE' n(E')$, this gives:
\begin{equation}
\delta N(E) = -\frac{1}{\pi} \, \left.\arg (G^{\,0}_{\bm{00}})\right]_{-\infty}^E \; .
\end{equation}
This is a typical phase shift formula which can be derived quite generally by starting from the identity:\cite{Lifshitz1965}
\begin{equation}
 N(E) =  \frac{\text{Im}}{\pi}\text{Tr} \ln G =  \frac{\text{Im}}{\pi}\ln \det G \; .
 \end{equation}
The behaviour of $\arg (G^{\,0}_{00})$ is particularly simple to study if we use an Argand plot in the plane (Re $ G^{\,0}_{00}, \text{Im} \,G^{\,0}_{00})$ and look at the trajectory of $G^{\,0}_{00}$ as a function of $E$. It is more convenient to work with $-G^{\,0}_{00}$ which has a positive imaginary part so that the angle $\theta = \arg(-G^{\,0}_{00})$ belongs to the interval $[0,\pi]$ and $\delta N(E) = -\theta/\pi$.  The usual definition for the phase shift $\eta$ is $\eta =
\arg(1/G^{\,0}_{00}) = -\theta$. It is clear from  Fig.\ \ref{arg} that since  $G^{\,0}_{00}$ vanishes at the origin, $\theta$ should jump from $\pi$ to zero. Actually, because of the weakness of the logarithmic singularity of the real part of the Green function, the jump is limited by the value of $\varepsilon$, here equal to 10$^{-4}$ and by the resolution of the figure. 

\begin{figure}[!fht]
\centering
\includegraphics[width=6cm]{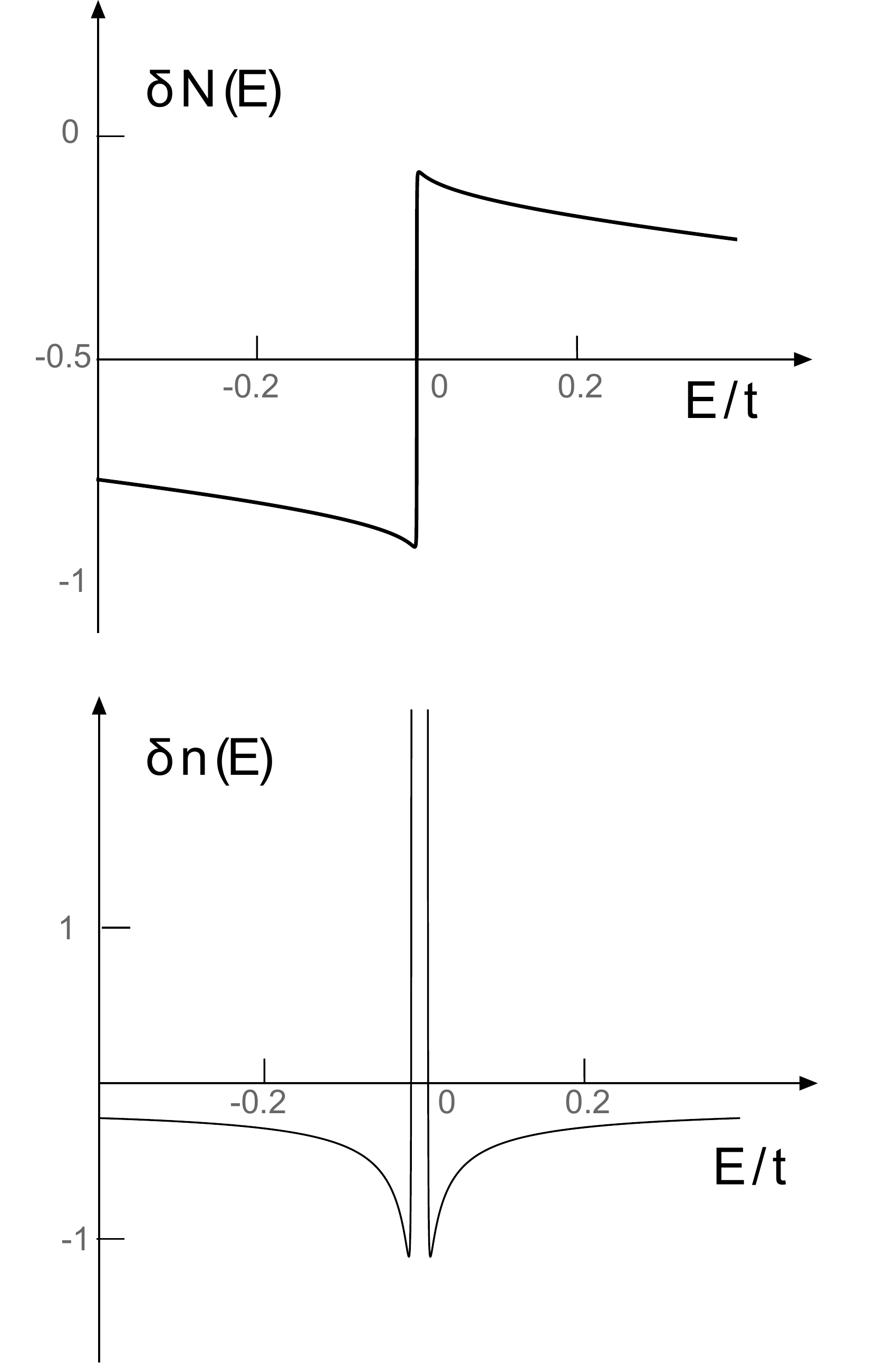}
\caption{Variations as a function of $E$  of  $\delta N(E) $(top) and of $\delta n(E)$ (bottom).}
\label{eta}
\end{figure}

The corresponding variations of $\delta N(E)$ and of its derivative $\delta n(E)$ are shown in Fig.\ \ref{eta}. The latter variation has therefore a $\delta$-function contribution of weight one and also a singularity close to $E=0$. This can be shown directly since $d/dz (G^{\,0}_{00}(z)) = 1/z + 2/(W^2 G^{\,0}_{00}(z) )$ so that, close to $E=0$:
\begin{equation}
 \delta n( E) \simeq \delta(E) -  \frac{2}{|E|\left(\ln(W^2/E^2\right)^2}  \; .
 \end{equation}
$\delta N(E)$ has also particular behaviours close to van Hove singularities which are not well reproduced by our model but which can be qualitatively understood using Argand plots. The striking point is that since $\delta N(E) \to -1$ when $E\to\infty$  two ``states'' are substracted from the initial system, one state when $E<0$ and another one when $E>0$. One state corresponds to the vacancy missing state (in the model with a local potential $U \to \infty$ this state 
has been pushed to infinity) and the other one to the $\delta$-function. What is surprising here is that this $\delta$-function \textit{does not} correspond to a single well-defined zero mode but is the result of interferences between many resonant states close to $E=0$. The fact that the $\delta$-function has weight one exactly is related to the fact that $G^{\,0}_{00}(z)$ vanishes at $z=0$ and that its real part has a logarithmic singularity (infinite slope). Actually as is clear in Fig.\ \ref{eta} the practical weight is lower than one as soon as we integrate in a small window around $E=0$.
Introducing a small imaginary part amounts in collision theory to consider wave packets of extension $\sim 1/\varepsilon$ (the Green functions decay exponentially on this scale), very large, but much smaller that the overall size of the system. The wave functions can then be calculated using the Lippmann-Schwinger equation as  noticed by Nanda \textit{et al.} \cite{Nanda2012} Although not valid strictly at $E=0$ it can be used in the limit $E \to 0$ to show that all states there are strongly hybridized with the zero mode  $G^0 |0\rangle$.

\subsubsection{Application: spatial variation of the electronic density}
\begin{figure}[!fht]
\centering
\includegraphics[width=6
cm]{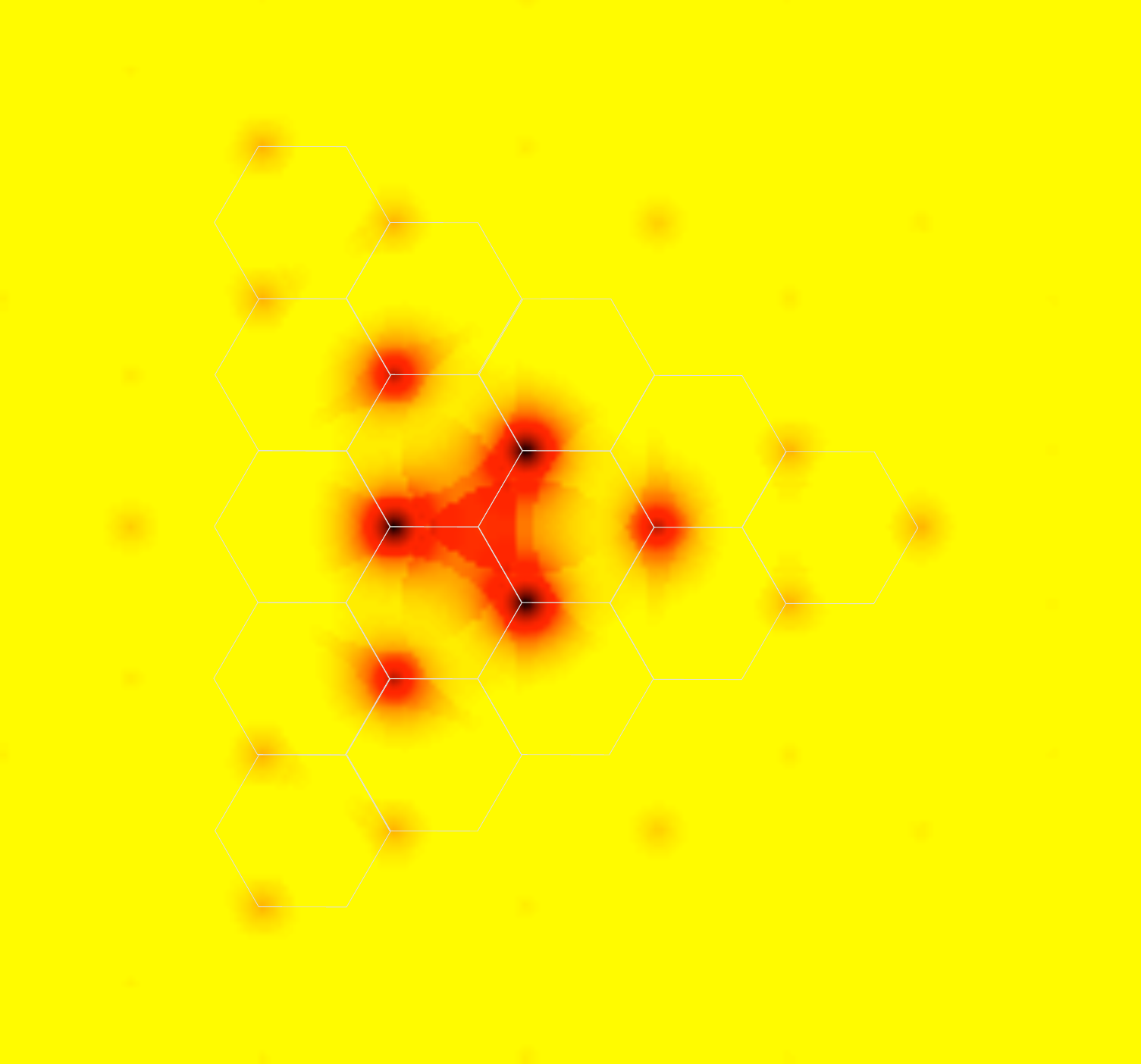}
\caption{Local density of states in the presence of a vacancy (courtesy of H.Amara).}
\label{imlac}
\end{figure}

The electronic density at point $\bm{r}$ is given by $n(\bm{r},E)=\langle\bm{r}|\,\delta(E-H)|\bm{r}\rangle $. Projecting on the $\pi$ atomic basis, we obtain:
\begin{eqnarray}
n(\bm{r},E)&= &\sum_{\bm{n,m}}  \langle\bm{n}|\,\delta(E-H)|\bm{m}\rangle
\phi(\bm{r}-\bm{n})\,\phi(\bm{r}-\bm{m}) \\
&=&
 -\lim_{\epsilon\to 0} \sum_{\bm{n,m}}  \frac{\text{Im}}{\pi} \,G_{\bm{nm}}(E+i\varepsilon) \phi(\bm{r}-\bm{n})\,\phi(\bm{r}-\bm{m})\, ,\nonumber
\end{eqnarray}
where $\phi(\bm{r}-\bm{n})$ is the $\pi$ orbital at site $\bm{n}$. The main contribution to the sum is provided by the diagonal terms but interference terms related to the overlap between neighbouring orbitals are not always negligible. In the case of the vacancy, we have an exact expression for $G_{\bm{nm}}(E)$ (see Eq.\ (\ref{Goff})). Close to zero energy the unperturbed electronic density vanishes so that the main contribution to $n(\bm{r},E)$ is the diagonal term proportional to $(G^{\,0AB}_{0\bm{n}}(E=0))^2$ (see Eq.\ (\ref{eldens})). As a result  the spatial variation of $n(\bm{r},E\simeq 0)$ is driven by the behaviour of the Green functions. As mentioned previously, in the ``Bessel'' approximation these functions contain a geometrical factor shown in Fig.\ \ref{GAB++}  and a spatial dependence $\sim 1/n $. The result is shown in Fig.\ \ref{imlac}. For clarity the atomc orbitals have just been replaced by s-like functions. The resulting electronic density then mimics what can be observed for example in STM observations at a finite height above the graphene plane. The vacancy (A site) is at the center of the central triangle. The electronic density is concentrated on B sites. The image is characterized by the presence of two bright ``triangles" and three arms formed by losanges. Three isolated dots can also be viewed as well as extinctions due to the geometrical factor $\cos  (\bm{K}.\bm{n}-\omega_{\bm{n}})$. This type of image is actually typical of that produced by a local resonant defect. \cite{Wehling2007,Farjam2011} From the experimental side such images have been observed in nitrogen doped samples.\cite{Zhao2011,Joucken2012} They can be due to substitutional nitrogen atoms and/or to so-called pyridine defects. They have also been clearly identified in STM observations. Notice that this typical contrast \textit{is not} what is frequently mentioned as a $\sqrt{3}\times\sqrt{3}$ contrast, which has been observed in carbon nanotubes \cite{Clauss1999, Buchs2007, Furukashi2008} as well as in graphene in the presence of extended defects.\cite{Sakai2010,Kim2013}\\

\subsubsection{Finite number of vacancies}
Vacancies on the same sublattice do not interact so that the difficult situations are those of ``compensated" systems where the number of vacancies is the same on both sublattices. The t-matrix becomes here a genuine matrix in the space defined by the set of states on the vacancy sites. The label $c$ will denote sites of the vacancy cluster $C$  and $P_C$ is the projector on these sites, $P_C= \sum_c |c\rangle\langle c|$. Introducing potentials $U \to \infty$ on each 
vacancy site, the t-matrix becomes  $-(P_CG^{\,0}P_C)^{-1}$. Consider first a pair of vacancies on two sites 1 and 2 belonging to sublattices $A$ and $B$. $P_CG^{\,0}P_C$ becomes a $2 \times 2$ matrix whose diagonal matrix elements are equal to $G^{\,0}_{00}$. In the limit $E \to 0$, the off-diagonal elements $G^{\,0AB}_{12}$ are real and independent of the energy, so that the resonance at $E=0$ of a single vacancy is replaced by a pair of bonding-antibonding states at energies given by:
\begin{equation}
\frac{E}{W^2} \ln \frac{E^2}{W^2} = ±\pm G^0_{12} \; .
\end{equation}
Notice here that the equation $x \ln| x| =y$ has the asymtotic solution $x=y/\ln| y| + \dots$. Actually for first neighbours, $G^0_{12}$ is not really small, the imaginary parts can no longer be neglected and the low energy limit is no longer satisfied. Resonances close to zero energy will only occur for large separations of the vacancies. Consider now a general cluster containing $N_v$ vacancies of each type. 
The condition for resonances becomes $\det (P_CG^{\,0}P_C)=0$. Close to zero energy the only non-vanishing elements of $P_CG^{\,0}P_C$ are the real part of the $G^{\,0AB}$ Green functions. We have therefore to study the eigenvalues of $P_CG^{\,0}P_C$ considered (at zero energy) as an effective hamiltonian between vacancies on different sublattices, with long range $1/n$ interactions.\cite{Huang2010} This is obviously a difficult problem in the case of a finite concentration of vacancies similar to the impurity band problem. \cite{Lifshitz1965} 

To summarize, we have clarified some points related to the electronic structure of a finite number of vacancies in graphene. The case of a finite vacancy concentration will be dealt with elsewhere.

\acknowledgments
Numerous discussions with H. Amara are gratefully acknowledged.

\end{document}